\documentclass[prd,aps,floats,epsfig,eqsecnum,nofootinbib]{revtex4}
\usepackage{array,amsmath,amssymb,verbatim,epsfig,rotating}
\begin{document}
\title{\bf{Conceptual unification of elementary particles, black holes, 
quantum de Sitter and Anti de Sitter string states}}
\author{Norma G. SANCHEZ \\
Observatoire de Paris, LERMA \\61, avenue de l'Observatoire \\ 75014 Paris, 
FRANCE \\ Norma.Sanchez@obspm.fr,\quad wwwusr.obspm.fr/sanchez}
\date{\today }
\begin{abstract}
We provide a conceptual unified description of the quantum properties of
black holes (BH), elementary particles, de Sitter (dS) and Anti de Sitter
(AdS) string states.The conducting line of argument is the classical-quantum
(de Broglie, Compton) duality here extended to the quantum gravity (string) regime
(wave-particle-string duality). The semiclassical (QFT) and quantum (string)
gravity regimes are respectively characterized and related: sizes, masses, accelerations and
temperatures. The Hawking temperature, elementary particle and string
temperatures are shown to be the same concept in different energy regimes and turn out
the precise classical-quantum duals of each other; similarly, this result holds for the BH decay rate,
heavy particle and string decay rates; BH evaporation ends as quantum string decay into pure (non mixed) radiation. Microscopic density of states and
entropies in the two (semiclassical and quantum) gravity regimes are derived
and related, an unifying formula for BH, dS and AdS states is provided in the two regimes. A string phase transition
towards the dS string temperature (which is shown to be the precise quantum dual of the semiclassical (Hawking-Gibbons) dS temperature) 
is found and characterized; such phase transition does not occurs in AdS alone. High string masses
(temperatures) show a further (square root temperature behaviour) sector in AdS. From the string
mass spectrum and string density of states in curved backgrounds, quantum properties of the
backgrounds themselves are extracted and the quantum mass spectrum of BH,
dS and AdS radii obtained.
\end{abstract}
\maketitle
 \tableofcontents

\section{Introduction and results}
Macroscopic black holes arise through the gravitational collapse of stellar 
bodies. Microscopic black holes could arise from high density concentrations 
(of the order of the Planck energy scale) in the early universe, as well as 
from the collisions of particles at such energy scales. Microscopic black 
holes are necessarily quantum and their properties governed by quantum or 
semiclassical gravity, evaporation through Hawking radiation is a typical 
effect of these black holes. Microscopic black holes share in some respects 
analogies with elementary particles, and on the other hand, show many 
important differences. A theory of quantum gravity, or ``theory of
everything'' such as string theory, should accounts for an unified and
consistent  
description of both black holes and elementary particles, and the physics of 
the early universe as well. \\
In this paper, we provide an unifying description of the quantum properties 
of black holes, elementary particles, de Sitter and Anti-de Sitter states. 
The conducting line of argument is the concept of classical-quantum 
(wave-particle) duality at the basis of quantum mechanics, here extended 
to include the quantum gravity string domain, ie wave-particle-string 
duality. We set up the relevant scales characteristic of the semiclassical 
gravity regime (QFT matter + classical gravity) and relate it to the 
classical and quantum gravity regimes. The de Broglie-Compton wave length $L_q=\hbar (M_{cl}c)^{-1}$ 
in the presence of gravity means  
$L_q=l_{Pl}^2\, L_{cl}^{-1}$ (``Planck'' duality), $L_{cl}$ and $l_{Pl}$ 
being the gravitational ($L_{cl}=\frac{G}{c^2}\, M_{cl}$) and Planck length scales 
respectively. As quantum behaviour is the dual of classical behaviour 
(through $\hbar ,\, c$), the dual of semiclassical gravity regime is full 
quantum gravity regime (through $\hbar ,\, G,\, c$), (in string theory 
through $\hbar ,\, \alpha ',\, c$). The semiclassical gravity mass scale 
is the ``dual'' mass $M_{sem}=m_{Pl}^2\, M_{cl}^{-1}$, its corresponding 
temperature scale (energy) 
is the Hawking temperature $T_{sem}=\frac{1}{2\pi k_B}\, M_{sem}\, c^2$, 
($m_{Pl} \equiv $ Planck mass).\\
The set of quantities $O_{cl,\, sem}=(L_{cl},\, M_{cl},\, \mathcal{K}_{cl},\, 
T_{sem})$ characteristic of the classical/semiclassical gravity regime, 
(here denoting size, mass, gravity acceleration, Hawking temperature 
respectively), and the corresponding set in the full quantum gravity 
regime $O_q=(L_q,\, M_{q},\, \mathcal{K}_q,\, T_q)$ are {\bf dual} (in the 
sense of the wave-particle duality) of each other, ie
\begin{equation}
O_{cl,\, sem}=o_{Pl}^2\, O_q^{-1}
\end{equation}
This relation holding for each quantity in the set. $O_{cl,\, sem}$ and 
$O_q$ being the same conceptual physical quantities in the different 
(classical/semiclassical and quantum) gravity regimes. ($o_{Pl}$ purely 
depending of ($\hbar ,\, c,\, G)$). $O_q$ standing for the usual concepts 
of quantum size, mass, acceleration and temperature (ie 
$T_q=\frac{1}{2\pi k_B}\, Mc^2$). In string theory we have similarly, 
\begin{equation}
O_{cl,\, sem}=o_s^2\, O_s^{-1}
\end{equation}
with $\alpha '$ instead of $G/c^2$ (and $o_s^2$ purely depending on 
($\hbar ,\, \alpha ',\, c$)). $O_s=(L_s,\, M_s,\, \mathcal{K}_s,\, T_s)$ 
denoting here the characteristic string size, string mass, string 
acceleration and string temperature of the system under consideration, 
($T_s=\frac{1}{2\pi k_B}\, M_s\, c^2$), which are in general different from 
the usual flat space string expressions, (in particular they can be equal). 
This is not an assumed or conjectured duality: the results of QFT and 
string quantization in black holes, de Sitter, Anti-de Sitter and WZWN 
backgrounds remarkably show these relations \cite{1},\cite{2},\cite{3} and sections below. As the wave-particle duality, 
the semiclassical-quantum gravity duality does not relate to the number of 
dimensions, nor to a particular symmetry. In this paper we also extend and 
use these relations to include the microscopic density of states and 
entropies ($\rho _{sem},\, S_{sem}$) and ($\rho _s,\, S_s$), as well. The 
gravitational size $L_{cl}$ in the classical regime becomes the quantum 
size $L_q$ or string size $L_s$ in the quantum gravity regime. The 
QFT Hawking temperature $T_{sem}$ in the semiclassical gravity regime 
becomes the string temperature $T_s$ in the full quantum gravity regime.
We find ($\rho _{sem},\, S_{sem}$) in the semiclassical gravity regime such 
that they become ($\rho _s,\, S_s$) in the quantum gravity string regime. 
The asymptotic (high M) expressions are given by 
\begin{eqnarray}
&&e^{S_{sem}/k_B}=\rho _{sem}(M_{cl})=\left( \frac{S_{sem}^{(0)}}{k_B}\right) ^{-a} 
e^{S_{sem}^{(0)}/k_B} \; , \quad \mbox{where}  \quad \cr \cr
&&S_{sem}^{(0)}=\frac{k_B}{4}\, \frac{A}{l_{Pl}^2}=\frac{\pi k_B}{p}\, 
\frac{T(M_{cl})}{T_{sem}}=\frac{1}{2p}\, \frac{M_{cl}c^2}{T_{sem}}
\end{eqnarray}
(here $a=D$ space time dimensions, and with our normalization of $L_{cl}=\frac{G}{c^2}\, M_{cl}$ : $p=4$ for BH's, $p=1$ for dS and AdS). 
The leading term $S_{sem}^{(0)}$ is the known gravitational entropy ($A\equiv$ horizon area), 
(in the absence of event horizon, as in AdS, A is the area associated to 
$L_{cl}$, as $T_{sem}$ just becomes a temperature scale associated to $M_{sem}$, 
the Hawking temperature in AdS being formally zero). 
\\
The string density of mass states and string
entropy in $BH$ (asymptotically flat) backgrounds are the same as in 
flat space time \cite{4}. We find $\rho _s$ and $S_s$ in dS and AdS backgrounds which are different 
from their flat space expressions (the string mass spectra in dS and AdS are different from the flat space expressions) \cite{5},\cite{6},\cite{7}). 
The asymptotic formulae (high M) are 
\begin{equation}
e^{S_s/k_B}=\rho _s(M)=f(M)\left( \frac{S_s^{(0)}}{k_B}\right) ^{-a} 
e^{S_s^{(0)}/k_B}
\end{equation}
\[
S_S ^{(0)}(M)=\frac{1}{2\pi }\frac{Mc^2}{T_S },
\quad
T_S =\frac{1}{2\pi k_B }M_S c^2
\]
where,
\begin{eqnarray}
&&p= 4,  \quad M_S =\frac{1}{8b}\sqrt {\frac{\hbar }{\alpha 'c}} ,\quad
f=1:\,\quad BH \cr \cr
&& p= 1, \quad M_S =\frac{1}{8b}\frac{c}{\alpha 'H}, \quad f=\sqrt {\frac{M_S }{M_S -M}} :\quad dS \cr \cr
&& p=1,\quad M_S =\frac{1}{8b}\frac{c}{\alpha 'H},\quad f=1:\quad AdS \; .
\end{eqnarray}
Here H is the Hubble constant and $b=2\sqrt{\frac{D-2}{6}}$, (D = number of space-time dimensions). In BH's and AdS backgrounds, high string masses are not bounded, while in dS
space-time $M_S $ is a maximal mass for the oscillating (particle) string
states. Remarkably, for $M\to M_S $, the string density of states and
entropy $(\rho _S ,S_S )$ indicate the presence of a phase transition at the
de Sitter string temperature $T_S $; the square root branch behaviour near
$T_S $ is universal: it holds in any number of dimensions, and is analogous to
that found in the thermal self-gravitating gas of (non-relativistic) particles (by mean field
and Monte Carlo methods) \cite{8}. In AdS background alone (including WZWN
models) such phase transition does not occur, (there is no finite (at finite temperature) critical point: (the negative $\Lambda $ (curvature)
pushes it to infinity ; $(\rho _S ,S_S )$, the partition function are all finite in AdS). (The Hawking-Page
transition \cite{9} which occurs in the context of semiclassical BH-AdS systems is due
to the BH, not to AdS which acts only as a space boundary condition). In AdS
backgrounds (including WZWN models) the very high string masses ($M>>M_S$) show a
new sector (non existing in BH's nor in dS background):
\[
S_S ^{(0)}(M>>M_S )=\pi k_B \frac{\sqrt {MM_S } }{m_S }=\frac{\sqrt {TT_S }
}{t_S }
\]
The characteristic mass ratio (or temperature) in $\rho _S (M)$ and $S_S (M)$
from the low to the high masses follows the behaviour
\[
\quad
(i)\frac{M}{m_S }(M<<M_S )\to (ii)\frac{M_S }{m_S }(M\approx M_S )\to
(iii)\frac{\sqrt {MM_S } }{m_S }(M>>M_S) ,
\]
which, with our dual relation $\frac{t_S }{T_S }=\frac{T_{sem} }{t_S },$ reads
\[
(i)\frac{T}{t_S }(TT_{sem} <<t_S ^2)\to (ii)\frac{t_S }{T_{sem} }(TT_{sem}
\approx t_S ^2)\to (iii)\sqrt {\frac{T}{T_{sem} }} (TT_{sem} >>t_S ^2)
\]
The sector (iii) is absent in BH's and dS backgrounds. Interestingly enough,
the highly excited string spectrum and our classical-quantum gravity dual relations allow to infer quantum properties of the
background itself. The mass scale for the low masses is the fundamental
string mass $m_S =\sqrt {\hbar /c\alpha '} $ while for the high masses the
scale is $M_S $. For $m\to M_S $, the string becomes the background (and
conversely, the background becomes the string): it turns out that $M_S $ is
the mass $M_{cl}$ of the background for $\left( {\raise0.7ex\hbox{$G$}
\!\mathord{\left/ {\vphantom {G
{c^2}}}\right.\kern-\nulldelimiterspace}\!\lower0.7ex\hbox{${c^2}$}}
\right)\to \alpha '$, and conversely (ie $M_{cl}\Leftrightarrow M_S$ with$\left(
{\raise0.7ex\hbox{$G$} \!\mathord{\left/ {\vphantom {G
{c^2}}}\right.\kern-\nulldelimiterspace}\!\lower0.7ex\hbox{${c^2}$}}
\right)\Leftrightarrow \alpha ')$. The massive string properties reproduce
those of the background, and the quantum mass spectrum of the background is
obtained such that it becomes the highly massive string spectrum in the
quantum gravity regime. We obtain
\[
M_{cl}=(\frac{2\pi k_B }{c^2}T_{sem} )2pb^2N,\quad ie\quad M_{cl}=m_{mpl} 2pb\sqrt N ,
(b=2\sqrt {\frac{D-2}{6}} )
\]
where p=4 for BH's, p=1 for dS and AdS. This spectrum hold in any number of dimensions. This is like the high N mass
spectrum of strings in BH's, dS and AdS backgrounds when $M$ tends to $M_S$. For the BH and dS
radii (and AdS scale length) we have 
\[
R_{BH} =\frac{(D-3)}{2}l_{pl} 8b\sqrt N ,
\quad
\frac{c}{H}=2bl_{pl} \sqrt N
\]
Quantum string properties of BH's, dS and AdS states are determined by $(L_S
,M_S ,K_S ,T_S )$ (size, mass, acceleration and temperature of strings in
these backgrounds), and their quantum spectra determined by the highly
excited string spectra. These string phase properties confirm and complete
the back reaction results found in the string analogue model
(thermodynamical approach)\cite{1},\cite{2}). The QFT semiclassical BH's, dS and AdS
backgrounds are a low curvature (low energy) phase (for $T_{sem} <<T_S )$ of
the string phase reached when $T_{sem} \to T_S $, (and so when $L_{cl} \to L_S
,K_{cl} \to K_S )$. The two phases are dual (in the sense of the
wave-particle-string duality) of each other. From these findings, the last
phase of black hole evaporation is shown to follow the same decay formula as
heavy elementary particle decay or string decay. This completes in precise
way the BH evaporation picture computed in ref\cite{2}: string emission by BH's
is an incomplete gamma function of $(T_S -T_{sem} )$ which for $T_{sem} <<T_S
$ yields the QFT Hawking emission, and for $T_{sem} \to T_S $ undergoes
(back reaction) a phase transition to a string state which decays (as a
string) into (most massless) particles including the graviton. As shown here, BH evaporation
evolves precisely from a semiclassical gravity QFT phase (Hawking emission)
with decay rate $\Gamma _{sem} =\left| {{\dot {M}} \mathord{\left/
{\vphantom {{\dot {M}} M}} \right. \kern-\nulldelimiterspace} M}
\right|\approx GT_{sem} ^3$ to a quantum gravity string phase decaying as
pure (non mixed) radiation with decay rate $\Gamma _S =GT_S ^3 $. Conversely,
cosmological evolution goes from a quantum string phase (selfsustained by
strings with $T_{S} >> T_{sem}$) to a semiclassical QFT phase (QFT inflation) and then to the classical
(standard FRW) phase. The wave-particle-string duality precisely
manifests in this evolution, between the different gravity regimes, and can be
view as a mapping between asymptotic (in and out) states characterized by
the sets $O_{cl, sem} $ and $O_S $, and so as a 
S-matrix description. 
\\ 
This paper is organized as follows: in
Section II we elaborate on the concept of wave-particle-string duality.
Section III deals with semiclassical (QFT) and quantum (string) BH's: their
microscopic density of states, entropies, quantum BH spectrum and decay. In
sections IV and V we treat similar issues for the semiclassical and quantum
dS and AdS states respectively, derived from the mass spectrum, density of
string states and entropies. Some concluding remarks are given in section
VI.
\section{Wave-particle-string duality}
\subsection{Classical-quantum duality}
In classical physics in the absence of gravity, there is no relation between 
mass and length (there is no G), and there is no temperature associated to 
the length. 
From a classical length $L_{cl}$ one constructs an ``acceleration''  
$\mathcal{K} _{cl}$ 
(through c), and to a mass $M_{cl}$ one associates a temperature 
(through Boltzmann constant $k_B$) : 
\begin{equation}
L_{cl} \stackrel{c}{\rightarrow} \mathcal{K} _{cl} = \frac{c^2}{L_{cl}} \quad 
, \quad T_{cl} = \frac{1}{2\pi k_B} M_{cl}c^2\, .
\end{equation}
\\
In quantum physics (QFT), mass and length are related through $\hbar $ by 
the Compton 
length $L_q$. One   
constructs a (quantum) ``acceleration''  
$\mathcal{K} _q$ 
(through c), and an associated (quantum) temperature $T_q$ (through $k _B$) : 
\begin{equation}
\lambda _q \equiv L_q = \frac{\hbar }{mc}, \quad \quad \mathcal{K} _q = 
\frac{c^2}
{L_q},  
\quad \quad T_q = \frac{1}{2\pi k_B} mc^2,
\end{equation}
which can be also expressed as 
\begin{equation}
\mathcal{K} _q = \frac{c^3}{\hbar }m, \quad \quad T_q = \frac{\hbar c}
{2\pi k_B}L_q ^{-1} 
= \frac{\hbar }{2\pi k_Bc}\mathcal{K} _q \, .
\end{equation}
In the presence of gravity, length and mass are related 
(through G), the classical line eq (2.1) is written as : 
\begin{equation}
L_{cl} = \frac{G}{c^2}M_{cl}, \quad \mathcal{K} _{cl} = 
\frac{c^4}{G}M_{cl}^{-1}, 
\quad T_{cl} = \frac{1}{2\pi k_B}\, \frac{c^4}{G}L_{cl}
\end{equation}
The quantum line eq (2.2) is thus related to $L_{cl}$   
through the Planck length $\sqrt{\hbar G/c^3} = l_{Pl}$ :   
\begin{equation}
L_q = \frac{l_{Pl}^2}{L_{cl}}, \quad \mathcal{K} _q = 
\frac{c^2}{l_{Pl}^2}L_{cl}, 
\quad T_q = \frac{\hbar c}{2\pi k_B} \,\frac{1}{l_{Pl}^2} L_{cl}  
 = T_{cl} 
\end{equation}
which also reads
\begin{equation}
L_q = \left( \frac{l_{Pl}}{c} \right) ^2 \, \mathcal{K} _{cl}^{-1}, \quad
\mathcal{K} _q = \left( \frac{c^2}{l_{Pl}} \right) ^2 \: 
\mathcal{K}  _{cl}^{-1}, \quad 
T_q = \frac{\hbar }{2\pi k_B}\left( \frac{c^2}{l_{Pl}}\right) ^2  
\mathcal{K} _{cl}^{-1}   
\end{equation}
(in terms of the classical ``acceleration'' $\mathcal{K} _{cl}$).  
Classical and quantum {\it lengths} are inversely related to each other 
through G, $\hbar $ and c. (Without G, such relation does 
not exist : $\hbar $ alone relates mass and quantum length  
, but not classical and quantum lengths). 
\subsection{Semiclassical-quantum gravity duality (QFT)}
In the context of QFT plus classical gravity, (semiclassical 
gravity), quantum matter is characterized by the  
relations eqs (2.2)-(2.3), classical gravity by the relations eq (2.4), and the 
combination of both gives rise to 
a non zero {\it semiclassical}  
temperature, which in the presence of event horizons is the Hawking temperature \cite{10} : 
\begin{equation}
T_{sem} = \frac{\hbar }{2\pi k_Bc}\, \mathcal{K} _{cl}
\end{equation}
$\mathcal{K} _{cl}$ is the classical acceleration eq.(2.1),(or surface gravity of the 
horizon) which in the black hole case is 
\begin{equation}
\mathcal{K} _{cl} = \frac{c^2}{L_{cl}}, \quad L_{cl} = \frac{2}{D-3}R_{BH}
\end{equation}
(D being the number of space time dimensions), $R_{BH}$ is the 
horizon radius. Therefore, the Hawking temperature which is a {\it semiclassical} concept 
for the black hole, can be expressed as  
\begin{equation}
T_{sem} = \frac{\hbar c}{2\pi k_B}L^{-1}_{cl}  
= \frac{\hbar c}{2\pi k_B}\, \frac{L_q}{l_{Pl}^2} 
= \frac{1}{2\pi k_B}\, m_{Pl}^2\, c^2\, M_{cl}^{-1}
\end{equation}
and we also write 
\begin{equation}
T_{sem}=\frac{1}{2\pi k_B}\: M_{sem}c^2, \quad 
M_{sem}=\frac{m_{Pl}^2}{M_{cl}}
\end{equation}
These expressions in terms of ($L_{cl}$, $K_{cl}$, $M_{cl}$) or 
($L_q$, $K_q$, $M_{sem}$) hold in any number of space-time dimensions, D 
enters in eq (2.8) and in the relation between $R_{BH}$ and the black 
hole mass $M_{BH}$ :
\begin{equation}
R_{BH} = \left( \frac{16\pi GM_{BH}}{c^2(D-2)A_{D-2}}\right)^{1/{(D-3)}} ,\quad 
\left( A_{D-2} \equiv \frac{2\pi ^{(D-1)/2}}{\Gamma ( \frac{D-1}
{2})} \right)  
\end{equation}
and so for the black hole  
\begin{equation}
M_{cl} = \frac{2}{(D-3)} \: \frac{c^2}{G} \: R_{BH} = 
\frac{2}{(D-3)}\: \frac{c^2}{G}\left( \frac{16\pi G M_{BH}}
{c^2(D-2)A _{D-2}}\right) ^{1/D-3}
\end{equation}
The Hawking temperature is a measure of the Compton length of the black hole, 
and thus of its quantum properties in the {\it semiclassical} (or QFT) 
regime, that is when the Compton length $L_q$ of the black hole is 
$l_{Pl}\ll L_q \ll L_{cl}$. 
Planck mass and Planck length satisfy by definition $
l_{Pl} = \frac{\hbar }{c}\, m_{Pl}^{-1}\quad \mbox{and}\quad 
l_{Pl} = \left( \frac{G}{c^2}\right) \,  m_{Pl}$.   
Classical and quantum black hole domains are related through the Planck 
scale : 
\begin{equation}
L_{cl}\: L_q = l_{Pl}^2 \quad, M_{cl}\: M_{sem} = m_{Pl}^2 \quad, 
\mathcal{K} _{cl}\: \mathcal{K} _q = \kappa _{Pl}^2 \quad, T_{sem}\:
T_q = t_{Pl}^2
\end{equation}
\begin{equation}
(l_{Pl}^2 = \hbar G/c^3 \quad,  m_{Pl}^2 = \hbar c/G
 \quad, \kappa _{Pl}\equiv  \frac{c^2}
{l_{Pl}},\quad  t_{Pl}\equiv \frac{1}{2\pi k_B}\, m_{Pl}c^2)
\end{equation}
\\ 
$\kappa _{Pl}$ and $t_{Pl}$ being the acceleration 
(or ``surface gravity'') 
and temperature at the Planck scale respectively.
$L_{cl}$, $\mathcal{K}_{cl}$, $L_{q}$, $\mathcal{K}_{q}$ are given by eqs (2.4), (2.5) respectively. 
Expressions (2.4), (2.5), (2.7) also yield
\begin{equation}
L_{cl}=\left( \frac{M}{m_{Pl}}\right) ^2\: L_q,\quad 
\mathcal{K}_{cl}=\left( \frac{m_{Pl}}{M}\right) ^2\: \mathcal{K}_q,\quad 
T_{sem}=\left( \frac{m_{Pl}}{M}\right)^2 \: T_q,
\end{equation} 
showing the change of regime through the Planck mass domain. The tension 
(mass/length) being $c^2/G$. 
\\\\
It must be noticed that eqs (2.13) not only hold for black holes, but more generally they relate the semiclassical and quantum regimes of gravity. The semiclassical mass scale, $M_{sem}$ eq. (2.10) we have introduced and its associated temperature scale eq. (2.10) are the characteristic scales of the semiclassical gravity regime. (In particular, in the presence of event horizons, $\mathcal {K}_{cl}$ is the surface gravity of the horizon and
$T_{sem}$ is the Hawking temperature).
\subsection{Semiclassical (QFT)-string quantum gravity duality.}
Quantum gravity string theory is naturally valid at 
the Planck scale. The Compton length of a quantum string is 
equal to its size 
\begin{equation}
l_s = L_q  = \frac{\hbar }{m_sc}\, ,
\end{equation}
and it also satisfies the ``gravitational''  
(length-mass) relation  
\begin{equation}
l_s = \alpha '\, m_s
\end{equation}
The fundamental string constant $\alpha '$ allows 
$l_s$ being {\it both} proportional and 
inversely proportional to $m_s$ :    
($\alpha '$playing the role of $\left[ \frac{G}{c^2} \right] $).In general, the two lentghs, $L_{q}$ and $l_{s}$ differ by a dimensionless number (string excitation), which we take here equal one (the ground state. 
All the relations eqs (2.2), (2.3) for  
elementary particles (QFT) hold universally for quantum strings (purely 
quantum objects), $(q\equiv s)$ : 
\begin{equation}
l_s = \frac{\hbar }{m_sc}\quad,\quad  \kappa _s = \frac{c^2}{l_s} \quad,
\quad t_s = \frac{\hbar c}{2\pi k_B}\,\frac{1}{l_s}
\end{equation}
$\kappa _s$, $t_s$ being respectively the fundamental string acceleration and 
string temperature 
\begin{equation}
\kappa _s = \frac{c^2}{\alpha ' \, m_s} = \frac{c^3}{\hbar }\, m_s, \quad \quad
t_s = \frac{\hbar }{2\pi k_Bc}\kappa _s = \frac{1}{2\pi k_B}\, m_sc^2
\end{equation}
\begin{equation}
(l_s = \sqrt {\frac{\hbar \alpha '}{c}} \quad, \quad 
m_s = \sqrt {\frac{\hbar }{c\alpha '}} \quad, \quad 
t_s = \frac{\hbar c}{2\pi k_B\alpha '} \, m_s^{-1})
\end{equation}
Eqs (2.15)-(2.16) can be expressed entirely in terms of $\alpha '$ 
(and $\hbar $, c), (instead of $l_{Pl}^2$),  
\begin{equation}
L_{cl}\,L_s = l_s^2,\quad 
M_{sem}\,M_s = m_s^2\: ,\quad 
\mathcal{K} _{cl}\, \mathcal{K} _s = \kappa _s^2 \: ,\quad 
T_{sem}\,T_s = t_s^2
\end{equation}
ie : 
\begin{equation}
L_s = \frac{\hbar \alpha '}{c}\,L_{cl}^{-1}, \quad 
M_s = \frac{\hbar }{c\alpha '}\,M_{sem}^{-1} \quad , \quad 
\mathcal{K} _s = \frac{c}{\hbar \alpha '}\,c^4 \mathcal{K} _{cl}^{-1},\quad 
T_s = \left( \frac{c}{2\pi k_B} \right) ^2 \frac{\hbar c}{\alpha '} \,
T_{sem}^{-1}
\end{equation}
Eqs (2.7)-(2.10) are the semiclassical expressions of eqs (2.21)-(2.22). In the 
quantum string regime, the Hawking temperature $T_{sem}$ 
becomes the string temperature $T_s$ (the black 
hole becomes a string)\cite{2},\cite{3}. Eqs (2.21)-(2.22) are the 
quantum 
string analogue of eqs (2.7)-(2.10).
The set ($L_s$, $M_s$, $K_s$, $T_s$) is the {\bf quantum 
string dual} (in the sense of the wave-particle string duality) of the 
classical/semiclassical (QFT) set ($L_{cl}$, $M_{sem}$, $K_{cl}$, $T_{sem}$). 
\\ \\
Eqs (2.20)-(2.21) are at the basis of the $\mathcal{R}$ transform 
introduced in refs\cite{1}-\cite{3}  in the context of Black 
holes and de Sitter space. Eqs (2.7)-(2.10) and (2.21)-(2.22) 
can be mapped one into another by a duality transform $\mathcal{R}$ 
\begin{equation}
\mathcal{R}[O_{cl}] = O_q = o_s^2 O_{cl}^{-1}\quad \mbox{or}\quad 
\mathcal{R}[O_{sem}] = O_s = o_s^2 O_{sem}^{-1}
\end{equation}
O being the same physical magnitude in the different classical, semiclassical 
and quantum or string regimes, $o_s$ purely depending on $\alpha '$ 
(and c, $\hbar $) (alternatively, $o_{Pl}$ 
purely depending on the Planck length : G, c, $\hbar $). 
An example of this $\mathcal{R}$ transform is the mapping of lengths 
$L_{cl}$ into  
$L_q$ (and all their related magnitudes), eqs (2.20)-(2.22). \\ \\
The dual of classical mechanics is quantum mechanics : a (wave) length is 
connected to the (inverse of) particle momentum through $\hbar $ (de 
Broglie 
wave length or ``de Broglie duality''). Its relativistic version (QFT),  
is the Compton length (``Compton duality''). 
In the presence of gravity,  
quantum and classical lengths are inversely connected to each other   
through the Planck length (``Planck duality''):   
semiclassical and quantum gravity regimes are dual of each other  
through the Planck scale domain.  
String theory being the consistent framework for 
quantum gravity, eqs (2.20)-(2.22) correspond 
to the quantum gravity string regime. As the dual of classical behaviour is quantum behaviour (with $\hbar $, c), 
the 
dual of semiclassical (QFT) gravity is full quantum (string) gravity (through
$\alpha '$, $\hbar $, c). This duality is {\it universal}. It is 
not linked to any symmetry, nor to the number or the kind of dimensions. \\ \\
The size of the black hole is the gravitational length $L_{cl}$ in the 
classical 
regime, it is the Compton length $L_q$ in the semiclassical regime, it is the 
string size $L_s$ in the full quantum gravity regime. Similarly, the horizon 
acceleration (surface gravity) $\mathcal{K}_{cl}$ of the black hole is 
the string acceleration $\mathcal{K}_s$ in the string regime. The  
Hawking temperature $T_{sem}$  
(measure of the surface gravity or of the 
Compton 
length) in the semiclassical gravity regime becomes the string temperature 
$T_s$ in the full quantum gravity 
(string) 
regime. Moreover, (see section 3 below), the mass spectrum of the 
black hole  in the 
semiclassical regime becomes the string spectrum in the full quantum regime. 
\\ \\
This duality does not needs {\it a priori} any symmetry nor compactified 
dimensions. It does not require the existence of any isometry in the curved 
background. 
Different 
types of relativistic quantum type operations 
$L \rightarrow L^{-1}$ appear in string theory due to the existence of the 
dimensional string constant $\alpha '$ (the most known being T-duality \cite{11},\cite{12},) linking 
physically equivalent string theories. The duality here we are considering 
is of the type classical-quantum (or wave-particle) duality (de Broglie ``
duality''), relating classical/semiclassical and quantum behaviours,  
here extended to include the quantum gravity string regime.\\ \\ 
The de Broglie 
and Compton wave-lengths $L_q = \hbar p^{-1}$ are not the expression 
of a symmetry transform between physically equivalent theories, but the 
crossing or the relationship between two different (classical and 
quantum) behaviours of Nature.  
The presence of G (or of $\alpha '$)  
yields for the Compton length  $L_q = l_{Pl}^2 \, L_{cl}^{-1}$, linking 
two different  
(semiclassical and quantum) gravity behaviours. This duality 
relation is supported explicitly from
the results of QFT and quantum strings  
in curved backgrounds \cite{1},\cite{2},\cite{3} and sections below. This is 
{\it not} an {\it assumed} or {\it conjectured} relation. 
The wave-particle-string duality, as the 
wave-particle duality at 
the basis of quantum mechanics, is reflected in the QFT and quantum string 
dynamics. As we will see below all relevant 
cases : black holes, de Sitter, AdS and WZWN-AdS backgrounds  
satisfy this QFT/string relation. The corresponding mass spectra and 
entropies as well (as we show in the sections below).
\section{Semiclassical (QFT) and Quantum (String) Black Holes}
Difficulties in connecting black holes to elementary particles 
appear when comparing the known black hole 
properties (which are semiclassical and thermal) to the elementary particle 
properties (which are quantum and intrinsically non-thermal). Even when 
comparing black holes and strings, most frequently there are the 
{\it semiclassical} black hole properties those compared 
to, or 
derived from, strings \cite {13}, \cite{14}. Semiclassical features (as Bekenstein- Hawking entropy) can be
derived in string theory, but the real issue here, comparison and 
unification of black holes and elementary particles, makes sense only for 
{\it quantum} black holes, and it takes place at the quantum gravity scale.\\ \\ 
The Hawking temperature  
is inversely proportional to the mass, while for an elementary particle,
temperature is 
equivalent (through units conversion) to the mass (energy). Thermal intrinsic properties of black holes as known till now from 
the QFT (or string) 
derivation are {\it semiclassical}. Microscopic black holes 
are {\it quantum} objects and the semiclassical treatment is not enough to disclose their connection to quantum elementary particles. Recall that the thermal 
black hole properties {\it are 
already present} (formally) at the {\it classical} level (the four laws of 
black hole mechanics)\cite{15}. Black hole thermodynamics  
can be compared to the thermodynamics of self gravitating (collapsed  
and relativistic) systems. In the quantum regime, quantum black holes become quantum strings, 
and the intrinsic thermal features of black holes are  
the intrinsic thermal features of strings. Quantum black holes at the 
Planck energy scale are just (become) particle (string) states. And/or, 
particle 
states at the Planck energy scale become quantum black holes. 
\subsection{Density of states and entropy}
The black hole density of mass states $\rho  (M)$ in the 
semiclassical (QFT) regime satisfy : 
\begin{equation}
\rho _{sem}(M) = e^{S_{sem}/k_B} \sim  e^{M_{cl}c^2/(8k_BT_{sem})}  
\end{equation}
where 
\begin{equation}
S_{sem} = \frac{k_Bc^3}{G\hbar} \,\frac{A_{BH}}{4} = 
\frac{\pi k_B}{4}\left( \frac{L_{cl}}{l_{Pl}} \right) ^2,  
\end{equation}
or,
\begin{equation}
S_{sem} = \frac{\pi k_B}{4} \left( \frac{M_{cl}}{m_{Pl}} \right) ^2 = 
\frac{1}{8}\: \frac{M_{cl}c^2}{T_{sem}}=\frac{\pi k_B}{4}
\: \frac{T_{cl}( M_{cl})}{T_{sem}}
\end{equation}
\\
$S_{sem}$ is the black hole entropy, $T_{cl}$ is the classical temperature eq (2.1) and $T_{sem}$ is 
the Hawking temperature eq (2.7) or (eqs (2.9), (2.10)). 
$L_{cl}$ and $M_{cl}$ are related to the black hole radius and mass by eqs (2.8) and (2.12) respectively 
(for D=4 : $L_{cl}=2R_{BH}$, $M_{cl}=4M_{BH}$).
The last two expressions in eq (3.3) are useful to compare with the 
corresponding  string quantities.  
$S_{sem}$ can be also expressed as  
\begin{equation}
S_{sem} = \frac{\pi k_B}{4} \left( \frac{M_{cl}}{M_{sem}}\right) 
 = \frac{\pi k_B}{4}\left( \frac{ L_{cl}}{L_q}\right) 
\end{equation}
and for comparison we also have 
\begin{equation}
S_{sem}\: S_s=s_{Pl}\, S_{QFT}, \quad \left( s_{pl}=\frac{\pi k_B}{4}\right)
\end{equation}
where
\begin{equation}
S_{QFT}=
\frac{\pi k_B}{4}\left( \frac{L}{l_{Pl}}\right) ^3 \quad \mbox{and}
\quad S_s=\frac{\pi k_B}{4}\left( \frac{L}{l_s}\right) 
\end{equation}

In pure QFT (without gravity) the number of modes of the fields is 
proportional to the 
volume of the system (ie a box), and a short distance external cut-off is 
necessary since QFT ultraviolet divergences, naturally placed here by 
quantum gravity scale $l_{Pl}$. The string entropy $S_s$ is proportional 
to the length. Clearly, the semiclassical gravity entropy $S_{sem}$ 
``interpolates'' between the pure (without gravity) QFT entropy $S_{QFT}$ and the 
(quantum gravity) string entropy $S_s$.
Expressions (3.3)-(3.5) for $S_{sem}$ explicitly exhibit its semiclassical 
nature, ie, $L_{cl}\gg l_{Pl}$ (equivalently, $M_{cl}\gg m_{Pl},\,
\mathcal{K} _{cl}\ll \kappa _{Pl},\, T_{sem}\ll t_{Pl}$). 
The expression
\begin{equation}
S_{sem} = \frac{k_B}{l_{Pl}^{D-2}}\: \frac{A_{BH}}{4}
\end{equation}
holds in any spacetime dimensions with the black hole area 
$A_{BH} = A_{D-2}\: R_{BH}^{D-2}$ 
($A_{D-2}$ and $R_{BH}$ being given by eq (2.11)). In terms of $L_{cl}, M_{cl}$, 
it reads :
\begin{equation}
S_{sem} = \frac{k_B}{4}\: A_{D-2}\left( \frac{(D-3)L_{cl}}{2l_{Pl}}\right) 
^{D-2} = 
\frac{\pi k_B}{4}\left( \frac{L_{clD}}{l_{Pl}}\right) ^2 = 
\frac{\pi k_B}{4}\left( \frac{M_{clD}}{m_{Pl}}\right) ^2
\end{equation}
with
\begin{equation} 
L_{clD}=\sqrt{\frac{A_{D-2}}{\pi }}\left( \frac{16\pi GM_{BH}}
{c^2(D-2)A_{D-2}}\right) ^{\frac{1}{2}\left( \frac{D-2}{D-3}\right) }
,\quad M_{clD} \equiv  \frac{c^2}{G}L_{clD}
\end{equation}
(for D=4 : $L_{cl4}=L_{cl}$, $M_{cl4}=M_{cl}$). Or,  
\begin{equation}
S_{sem}=\frac{1}{8}\left( \frac{M_{clD}c^2}{T_{sem}}\right) 
= \frac{\pi k_B}{4}\: \frac{T_{cl}(M_{clD})}{T_{sem}}
\end{equation}
\begin{equation}
T_{cl}=\frac{1}{2\pi k_B}M_{clD}c^2 \quad , \quad 
T_{sem} = \frac{1}{2\pi k_B}\, M_{semD}c^2
\end{equation}
\\
In the quantum gravity regime, black hole mass 
is quantized, several arguments in the context 
of QFT (horizon area quantization, canonical quantization) lead to the 
condition \cite{16},\cite{17},\cite{18} :
\begin{equation}
M_{cl}^2 = \frac{\hbar c}{G}N = m_{Pl}^2N, \quad N=0, 1,...
\quad \mbox{ie}\quad 
M_{cl} =  \left( \frac{2\pi k_B}{c^2}t_{Pl}\right) \sqrt{N}
\end{equation}
On the other hand, the string mass spectrum and string 
density of states in black hole (asymptotically flat)\cite{4} space times 
are the same as in flat space time, ie  
\begin{equation}
M^2 = \frac{\hbar }{c\alpha '}\, n = m^2_s n, \quad n = 0, 1, ...
\end{equation}
\begin{equation}
\rho _s(M) = e^{S_s/k_B} \sim e^{Mc^2/(8k_BT_s)} = e^{2\pi b \sqrt{n}}
\end{equation}
\begin{equation}
S_s = \frac{\pi k_B}{4} \,\left( \frac{L}{L_s} \right)  
= \frac{\pi k_B}{4} \left( \frac{ M}{M_s}\right)  = 
\frac{1}{8}\: \frac{M\, c^2}{T_s} 
\end{equation}
$S_s$ and $T_s$ being the string entropy and string temperature respectively,
\begin{equation}
S_s=2\pi k_B\: b\sqrt{n}, \quad T_s=\frac{t_s}{8b} = 
\frac{1}{2\pi k_B}\: \frac{m_sc^2}{8b}
\end{equation}
here
\begin{equation}
L_s=8b\, l_s, \quad M_s = \frac{m_s}{8b}
\end{equation}
ie,
\begin{equation}
M = \left( \frac{2\pi k_B}{c^2} \,t_s \right) \sqrt{n}=
\left( \frac{2\pi k_B}{c^2}\, 8b\, T_s\right) \sqrt{n}
\end{equation}
\\
$l_s,\: m_s$ being the fundamental string length and mass respectively eq (2.20).
The numerical coefficient b depends on D and on the type of strings, for 
instance :
\begin{equation} 
b=2\sqrt{\frac{D-2}{6}}\quad \mbox{for bosonic strings,
(closed and open)}. 
\end{equation}
Useful expressions for $T_s$ and $S_s$, (to compare to the black 
hole) are : 
\begin{equation}
T_s=\frac{\hbar c}{2\pi k_B}\: \frac{1}{L_s} = 
\frac{1}{2\pi k_B}M_sc^2 = 
\frac{c}{2\pi k_B}\: \frac{1}{8b}\sqrt{\frac{\hbar c}{\alpha '}}
\end{equation}
\begin{equation}
S_s=\frac{\pi k_B}{4}\: \frac{T(M)}{T_s}, \quad 
T(M)\equiv \frac{1}{2\pi k_B}Mc^2
\end{equation}
The black hole density of states and entropy ($\rho _{sem}$, $S_{sem}$) Eqs 
(3.1)-(3.3) are the semiclassical 
expressions of the string expressions ($\rho _s$, $S_s$) eqs (3.14)-(3.15). 
In the quantum string regime, $T_{sem}$ becomes  $T_s$, \cite{2}, \cite{3}.
[Eqs (3.1)-(3.4)] and [Eqs (3.14)-(3.15)] are the same 
physical quantities in different energy regimes. $\rho _{sem}$ and $\rho _s$ are given by eqs (3.1) and (3.14) respectively 
(at the leading behaviour).  
The black hole mass spectrum eq (3.12) becomes the string mass 
spectrum 
eq (3.18). We analyze it below.
\subsection{Quantum black hole spectrum}
Eq (3.14) is the leading term to $S_s$ for large M (large n). The asymptotic 
behaviour of the string degeneracy $d_n$ is
\begin{equation}
d_n \simeq (2\pi b\sqrt{n})^{-a}\: e^{2\pi b\sqrt{n}} =e^{S_s/k_B}
\end{equation}
Or, in terms of M 
\begin{equation}
e^{S_s/k_B}=\rho _s(M)\simeq \left( \frac{1}{8k_B}\: \frac{Mc^2}{T_s}\right) 
^{-a}\: e^{\frac{1}{8k_B}\: \frac{Mc^2}{T_s}}
\end{equation}
where for bosonic strings 
\begin{equation}
a = D \quad \mbox{(closed strings)}, \quad 
a = (D-1)/2 \quad \mbox{(open strings).}
\end{equation} 
From eq (3.23) and eqs (3.1)-(3.3) we are able 
to write the corrections to the black hole entropy $S_{sem}$, such that in 
the string limit $T_{sem}\rightarrow T_s$,
($\rho _{sem}$, $S_{sem}$) maps into ($\rho _s$, $S_s$) : 
\begin{equation}
e^{S_{sem}/k_B}=\rho _{sem}(M)\simeq \left( \frac{1}{8k_B}\: \frac{M_{cl}c^2}
{T_{sem}}\right) ^{-a}\: e^{\frac{1}{8k_B}\: \frac{M_{cl}c^2}{T_{sem}}}
\end{equation}
ie, 
\begin{equation}
S_{sem}=\frac{1}{8}\: \frac{M_{cl}c^2}{T_{sem}}-ak_B\log \left( \frac{1}{8k_B}
\: \frac{M_{cl}c^2}{T_{sem}}\right) 
\end{equation}
Or, from eq (3.3), (which we stand by $S_{sem}^{(0)}$, the leading term) : 
\begin{equation}
S_{sem}=S_{sem}^{(0)}-ak_B\log \left( \frac{S_{sem}^{(0)}}{k_B}\right) = 
\frac{\pi k_B}{4}\left( \frac{L_{cl}}{l_{Pl}}\right) ^2
- ak_B\log \left[ \frac{\pi }{4}\left( \frac{L_{cl}}{l_{Pl}}\right) 
^2\right] 
\end{equation}
This fix the logarithmic correction and its coefficient. Corrections to the semiclassical BH entropy have been continuously treated in the litterature, in the context of QFT \cite{38}, \cite{39},\cite{40} or strings (using correspondences, principles, ``strechted horizon'', ``Rindler wedge'', etc) \cite{14}, \cite{34}. We do not make use here of such assumptions or identifications.
\\ 
Eqs (3.25) and (3.23) leads us to the black hole quantization condition for the 
highly excited levels :   
\begin{equation}
\frac{1}{8k_B}\: \frac{M_{cl}c^2}{T_{sem}} = 2\pi 8b^2N, \quad \mbox{high N}
\end{equation} 
\begin{equation}
\mbox{ie} ,\quad M_{cl}=\left( \frac{2\pi k_B}{c^2}\: (8b)^2\: T_{sem}\right) N
\end{equation} 
This condition is such that in the quantum gravity regime,  
the black hole mass spectrum becomes the string mass spectrum. Eqs (3.28), (3.29) yields the black hole spectrum : 
\begin{equation}
S_{sem}=2\pi k_B\, 8b^2\, N - ak_B \log (2\pi 8b^2\, N)
\end{equation}
\begin{equation}
M_{cl}=m_{Pl}\, (8b\sqrt{N}),\quad
M_{sem}=\frac{m_{Pl}}{(8b\sqrt{N})}, \quad 
T_{sem}=\frac{t_{Pl}}{(8b\sqrt{N})}
\end{equation}
It displays the classical-quantum gravity dual nature eq(2.13),
$M_{sem}\, M_{cl} = m_{Pl}^2,\quad T_{sem}\, T_{cl}=t_{Pl}^2$. 
In any number of dimensions, the black hole mass spectrum eq (3.29) for $M_{cl}$ is like the flat space 
string spectrum eq (3.18) (for a string with mass $m_s=8bm_{Pl}$). $M_{cl}$ is connected to the black hole radius $R_{BH}$ and to the black hole 
mass $M_{BH}$ as given by eqs (2.11)-(2.12), we have then for $R_{BH}$ and 
$M_{BH}$ : 
 
\begin{equation}
R_{BH}=\frac{(D-3)}{2}\: l_{Pl}\: (8b\sqrt{N}),\quad 
M_{BH}=\mathcal{A}_D\left[ \frac{(D-3)}{2}\: l_{Pl}\: (8b\sqrt{N})
\right] ^{D-3},\quad\mathcal{A}_D\equiv \frac{c^2(D-2)}{16\pi G}\:A_{D-2}
\end{equation}

This is the quantized black hole radius and black hole mass in any number of dimensions.

\subsection{Unified quantum decay of QFT elementary particles, black holes and 
strings}
An unified description for the quantum decay rate of unstable heavy 
particles is provided by the formula\cite{19}
\begin{equation}
\Gamma = \frac{g^2\: m}{numerical\: factor}
\end{equation}
where g is the (dimensionless) coupling constant, m is the typical mass in 
the theory (the mass of the unstable particle object) and the numerical 
factor often contains the relevant mass ratios in the decay process. 
This formula nicely encompass all the particle width decays in the standard 
model (muons, Higgs, etc), as well as the decay width of topological and 
non topological solitons, cosmic defects and fundamental quantum strings \cite{19}. \\
A quantum closed string in an Nth excited state decays into lower 
excited states (including the dilaton, graviton and massless antisymmetric tensor fields) with a total width  
(to the dominant order (one string loop)) given by\cite{20}
\begin{equation}
\Gamma _s = \frac{G}{n^{\underline{o}}}T_s^3
\sim \frac{G}{l_s^3}
\end{equation}
which can be also written  
\begin{equation}
\Gamma _s = \frac{g^2}{n^{\underline{o}}}m_s,
\end{equation}
($n^{\underline{o}}$ is a numerical factor). 
That is, the string decay $\Gamma _s$ eq (3.35) 
has the same structure as 
eq (3.33) with $g \equiv \sqrt{\frac{G}{\alpha '}}$. \\ \\ 
On the other hand,  a semiclassical 
black hole decays 
thermally as a ``grey body'' 
at the Hawking temperature $T_{sem}$ (the ``grey body'' 
factor being the black 
hole absorption cross section $\sim L_{cl}^2$). The mass loss rate  
(estimated from a Stefan-Boltzmann relation) is
\begin{equation}
\left( \frac{dM_{cl}}{dt}\right)  = -\sigma L_{cl}^2T_{sem}^4 \sim 
T_{sem}^2
\end{equation}
where $\sigma $ is a constant. Then, the semiclassical black hole decay rate 
is given by 
\begin{equation}
\Gamma _{sem} = |\frac{d}{dt}\ln M_{cl} |  
\sim \frac{G}{n^{\underline{o}}}T_{sem}^3 \sim \frac{G}{L_{cl}
^3}
\end{equation}
As evaporation proceeds, the black hole temperature increases until it 
reaches the string temperature \cite{2}, the black hole enters its string regime 
$T_{sem} \rightarrow T_s ,\, L_{cl} \rightarrow L_s\: ,$ 
becomes a string state, then decays with a width 
\begin{equation}
\Gamma _{sem} \rightarrow GT_s^3 \sim \frac{G}{l_s^3}\rightarrow 
\Gamma _s 
\end{equation}
The semiclassical black hole decay rate $\Gamma _{sem}$ tends to the string 
decay rate $\Gamma _s$. 
Eq (3.27) is the semiclassical expression of eq (3.38). Again, unification between black hole decay and elementary particle decay 
is achieved for quantum black holes, when the black hole enters its 
quantum gravity (string) regime. 

\section{Semiclassical (QFT) and Quantum (String) de Sitter states}

The classical and semiclassical set of quantities ($L_{cl},\, 
\mathcal{K}_{cl}, 
\, M_{cl}, \, T_{sem}$) corresponding to  
QFT in de Sitter background \cite{21},\cite{22}, and the string set of quantities (
$L_s,\, \mathcal{K}_s,\, M_s,\, T_s$) derived from quantum string 
dynamics in de Sitter background \cite{23},\cite{5}, \cite{6}, \cite{2}(canonical as well as semiclassical quantization) are the following :
\begin{equation}
\mbox{QFT : }\quad L_{cl} = cH^{-1}, \quad  \mathcal{K}_{cl} = cH, \quad  
M_{cl} = \frac{c^3}{GH}
\end{equation}
\begin{equation}
T_{sem} = \frac{\hbar }{2\pi k_Bc}\, \mathcal{K}_{cl} = \frac{\hbar }{2\pi k_B}\, H = \frac{\hbar c}{2\pi k_B}L_{cl}^{-1}
\end{equation}
\begin{equation}
T_{sem}=\frac{1}{2\pi k_B}\: M_{sem}c^2, \quad M_{sem}=\frac{m_{Pl}^2}
{M_{cl}}=\frac{\hbar }{c^2}\: H
\end{equation}
\begin{equation}
\mbox{String : }\quad L_q \equiv L_s = \frac{\alpha '\hbar }{c^2}\, 
H = \frac{\alpha '\hbar }{cL_{cl}}
\end{equation}
\begin{equation}
\mathcal{K} _s = \frac{c^2}{L_s} = \frac{c^4}{\hbar \alpha ' H} = 
\frac{c^3}{\hbar \alpha '}\, L_{cl}
\end{equation}
\begin{equation}
M_s = \left( \frac{c}{\alpha 'H}\right) = \frac{\hbar }{cL_s} = 
\frac{1}{\alpha '}\, L_{cl}
\end{equation}
\begin{equation}
T_s = \frac{\hbar }{2\pi k_Bc}\, \mathcal{K} _s = \frac{c^3}{2\pi k_B}\, 
\frac{1}{\alpha ' H} = \frac{\hbar c}{2\pi k_B}\, \frac{1}{L_s}
 = \frac{1}{2\pi k_B}\, M_sc^2
\end{equation}
These expressions hold in any number of space-time dimensions (D): (D enters 
only through the relation between H and the scalar curvature, or cosmological 
constant $\Lambda $): 
\begin{equation}
R = D (D-1)\frac{H^2}{c}, \quad  H = c\: \sqrt{\frac{2\Lambda }{(D-1)(D-2)}}
\end{equation}
In the string quantities, D and the string model enter only through a 
numerical coefficient in $L_s$, as in flat space-time, (recall  
$L_{s\: flat} = \frac{8b(D-3)}{4\pi }\,l_s$, the numerical coefficient b 
depending on the string type (closed, open, supersymmetric, etc)). \\
As can be seen, the classical/semiclassical (QFT) set ($L_{cl},\,M_{sem},\,\mathcal{K}_{cl},
\,T_{sem}$) eqs (4.1)-(4.3) and the quantum string set ($L_s,\,M_s,\,
\mathcal{K}_s
,\,T_s$) eqs (4.4)-(4.7) satisfy the classical-quantum relations, eqs (2.21)-(2.22). 
The set ($L_s,\,M_s,\,\mathcal{K}_s,\,
T_s$) is the {\bf quantum string dual} (in the sense of the wave-particle-string 
duality) of the classical/QFT set ($L_{cl},\,M_{sem},\,\mathcal{K}_{cl},\,
T_{sem}$) :
\begin{equation}
L_s = l_s^2L_{cl}^{-1},\quad M_s = m_s^2M_{sem}^{-1},\quad 
\mathcal{K}_s = \kappa _s^2\mathcal{K}_{cl}^{-1}\: ,\quad 
T_s = t_s^2 T_{sem}^{-1}
\end{equation}
$L_{cl}$, $\mathcal{K}_{cl}$ and $M_{cl}$ are the radius, acceleration 
(surface gravity) and ``mass'' of the de Sitter (dS) universe respectively. 
$T_{sem}$ is the Hawking-Gibbons temperature: the QFT temperature in dS  
background, or the intrinsic dS temperature in its semiclassical(QFT) regime.  
$L_s$, $\mathcal{K}_s$ and $M_s$ are the size, acceleration and mass of 
quantum strings in dS universe. $T_s$ is the string temperature 
(the maximal temperature of strings) in dS background \cite{1},\cite{3}, which 
is also the intrinsic temperature of dS universe in the string regime : in the 
string analogue model with 
back reaction included, strings self-sustain a de Sitter phase of high 
curvature with temperature $T_s$ as given by eq (51) , \cite{1} and sub- sections below.
 
\subsection{Density of states and entropy} 

The entropy and density of states of semiclassical (QFT) de Sitter 
background are \cite{21},\cite{24}

\begin{equation}
\rho _{sem} = e^{S_{sem}/k_B}
\end{equation} 

\begin{equation}
S_{sem} = \frac{k_Bc^3}{G\hbar }\, \frac{A_{cl}}{4} = 
\pi k_B\left( \frac{L_{cl}}{l_{Pl}}\right) ^2 = 
\frac{\pi k_B}{l_{Pl}^2} \left( \frac{c}{H}\right) ^2
\end{equation}
Or,
\begin{equation}
S_{sem} = \pi k_B \left( \frac{M_{cl}}{m_{Pl}}\right) ^2 = \frac{1}{2}
\frac{M_{cl}c^2}{T_{sem}}=\pi k_B\frac{T_{cl}}{T_{sem}}
\end{equation}
On the other hand, the mass formula for quantum strings in the de Sitter 
background \cite{23},\cite{5},\cite{6} is given by :
\begin{equation}
\left( \frac{m}{m_s}\right) ^2 = 24\sum_{n>0}\: \frac{n^2+\omega _n^2(m)}
{\omega _n(m)}+2\mathcal{N}\: \frac{1+\omega _1^2(m)}{\omega _1(m)}
\end{equation}
\begin{equation}\mbox{where}~
\omega _n(m) = \sqrt{n^2-\left( \frac{m}{M_s}\right) ^2}\quad 
\mbox{and}\:  \mathcal{N}\:  \mbox{is the number operator}
\end{equation} 
\begin{equation}
\mathcal{N}=\frac{1}{2}\, \sum_{n>0}n\: [\alpha _n^{R+}\alpha _n^R + 
\tilde{\alpha} _n^{R+}\tilde{\alpha} _n^R],\quad  
[\alpha _n,\: \alpha _n^+]=1=[\tilde{\alpha}_n,\: \tilde{\alpha}_n^+]
\end{equation}
For large n : 
\begin{equation}
\left( \frac{m}{m_s}\right) ^2 \simeq 4n\left[ 1-n\left( \frac{m_s}{M_s}
\right) ^2 \right],\quad \mbox{with}\quad 
\left( \frac{m_s}{M_s}\right) ^2 = 
\frac{\alpha ' \hbar }{c}\left( \frac{H}{c}\right) ^2 = 
\left( \frac{l_s}{L_{cl}}\right) ^2 = \frac{L_s}{L_{cl}}
\end{equation}
The degeneracy $d_n(n)$ of level n (counting of oscillator states) is the same in flat as well as in curved space-time. 
The differences due to the background curvature enter through the relation $m=m(n)$ of the mass spectrum. The density $\rho(m)$ and the degeneracy $d_n$ satisfy the relation $\rho(m)dm= d_n(m)dn$. From this identity and eq.(4.13), we find  the string mass density of states  in de 
Sitter space as given by 
\begin{equation}
\rho _s (m) = f\left( \frac{m}{M_s}\right)  e^{2\pi b \frac{M_s}{m_s} 
\sqrt{1-\sqrt{1-\left( \frac{m}{M_s}\right) ^2}}}
\end{equation}
where  
\begin{equation}
f\left(\frac{m}{M_s}\right) = \frac{m/m_s}{\sqrt{1-\left( \frac{m}
{M_s}\right) ^2}}\: \frac{(m_s/M_s)^4}{\left( 1-\sqrt{1-\left( \frac{m}
{M_s}\right) ^2}\right) ^2}
\end{equation}
and $M_s$ is given by eq (4.6). For low ($m \ll M_s$) and high ($m\rightarrow M_s$) masses, we have 
respectively : 
\begin{equation}
\rho _s(m)_{m\ll M_s}  \sim \left( \frac{m}
{m_s}\right) ^{-a} e^{2\pi b\sqrt{\frac{
\alpha '\hbar }{c}}m \left[ 1-\frac{1}{8}\left( \frac{\alpha 'H}{c}\right) ^2 
m^2 + O\left( \frac{m}{M_s}\right) ^3\right] }
\end{equation}

\begin{equation}
\rho _s(m)_{m\rightarrow M_s} \sim \left( \frac{M_s}{m}\right) ^{-a}\left( 
\frac{M_s}{\Delta m}\right) ^{1/2} e^{2\pi b\frac{c}
{H}\sqrt{\frac{c}{\alpha '\hbar }}\: \left[ 1-\left( \frac{\Delta m}{2M_s}
\right) ^
{1/2}-\frac{1}{2}\left(\frac{\Delta m}{M_s}\right) + O\left( \frac{\Delta m}
{M_s}\right) ^{3/2} \right] }
\end{equation}
where a=D (closed strings) and 
\begin{equation}
\Delta m \equiv (M_s-m_s), \quad M_s = \frac{c}{\alpha ' H}
\end{equation}
\\ 
That is, (leading behaviour) : 
\begin{equation}
\rho _s(m)_{m\ll M_s} \sim \left( \frac{m}{m_s}\right)^{-a} e^{2\pi bm/m_s} 
\end{equation}
\begin{equation}
\rho _s(m)_{m\rightarrow M_s} \sim \left( \frac{M_s}{m_s}\right) ^{-a} 
\left( \frac{M_s}{\Delta m}\right) ^{1/2}
e^{2\pi bM_s/m_s} 
\end{equation}
Or, in terms of temperature : 
\begin{equation}
\rho (T\ll T_s)\simeq \left( \frac{T}{t_s}\right) ^{-a} e^{T/t_s} 
\end{equation}
\begin{equation}
\rho (T\rightarrow T_s)\simeq \left( \frac{T_s}{t_s}\right) ^{-a} 
\sqrt{\frac{T_s}{(T-T_s)}}\: e^{T_s/t_s}
\end{equation}
\begin{equation}
T \equiv \frac{1}{2\pi k_B}mc^2,\quad T_s\equiv \frac{1}
{2\pi k_B}M_sc^2=\frac{c^2}{2\pi k_B}\: (\alpha ' H)^{-1}
\end{equation}
For low masses, the string spectrum in dS background is like the  
flat space string spectrum \cite{1},\cite{5},\cite{6}. 
For high masses, the spectrum is saturated by the maximal mass 
$c/\alpha ' H$ for the oscillating (particle) states. The density   
$\rho _s(m)$ eqs (4.22 )-(4.23 ),  
similarly expresses in terms of the typical mass scale in each domain: 
$m_s$ (as in flat space) for low masses, $M_s$ for high 
masses corresponding to the 
maximal number of oscillating states : 
\begin{equation}
N_{max} \sim \mbox{Int} \left[ \frac{L_{cl}}{L_s}
\right]  
\sim \mbox{Int} 
\left[ \frac{c^3}{\hbar \alpha ' H^2}\right].   
\end{equation}  
(de Sitter space-time being bounded, this number although large is 
{\it finite}, and so arbitrary high masses can not be supported \cite{5},\cite{6},\cite{1},\cite{23}).
When $m > M_s$, the string does not oscillate  
(it inflates with the background, the  
proper string size is larger than the horizon \cite{25}), thus the string becomes ``{\it classical}\, '' reflecting the classical 
properties of the background. 
When $m \rightarrow M_s$, the string mass becomes the mass $M_{cl}$ of de Sitter 
background eq.(4.1), (with $\alpha '$ instead of $G/c^2$), and conversely, 
$M_s$ is also the mass of de Sitter background in its string 
regime, $M_s \Leftrightarrow (M_{cl})$. String back reaction in the 
string analogue model supports 
this fact : a de Sitter phase having  
mass $M_s$ and temperature $T_s$ given by eqs (4.6) and (4.7) respectively, is 
sustained by strings \cite{1}. Therefore,  
we interpret ($L_s$, $K_s$, $M_s$, $T_s$) given by eqs (4.4)-(4.7) as the 
{\it intrinsic} size, surface gravity , mass and temperature {\it of} 
de Sitter background in its string (high energy) regime. 
When the string mass becomes $M_s$, it saturates de Sitter universe (the 
string size $L_s$ (or Compton length) for this mass becomes the horizon 
size
$(\mbox{ie,}\: L_s \equiv L_q  = \frac{c}{H})$, that is the 
string 
becomes ``{\it classical} '', (or the background becomes quantum).  
The highly excited string states are very {\it quantum} as 
elementary particle states, but are ``{\it classical} '' as gravity states.    
The wave-particle-string duality  of quantum gravity reflects for the same 
states the 
classical gravity properties (with $\alpha '$ playing the role of $G/c^2$), 
and 
the highly quantum properties (with $L_s$ being the Compton length). \\ \\
De Sitter background is an exact solution of the semiclassical Einstein 
equations with the QFT back reaction of matter fields included \cite{22},\cite{26}. dS background is also a solution of the semiclassical Einstein equations with the string 
back reaction (in the string analogue model) included \cite{1}:  the 
curvature is a function of ($T_{sem}$, $T_s$) and contains the QFT 
semiclassical curvature as a particular case (for $T_{sem} \ll T_s$). The QFT 
semiclassical de Sitter background is a low energy  phase 
(for $T_{sem} < T_s$, 
$\mathcal{K}_{cl} < \mathcal{K}_s$). When $T_{sem}\rightarrow 
T_s$, then $\mathcal{K}_{cl} \rightarrow \mathcal{K}_s$ and from eqs (4.1) 
(4.4), (4.10) and (4.17 it follows that
\begin{equation}
L_{cl}\rightarrow L_s, \quad \mbox{ie} \quad \frac{c}{H} \rightarrow l_s ,
\end{equation}
the QFT semiclassical de Sitter phase becomes a string state selfsustained 
by a string cosmological constant
\begin{equation}
\Lambda _s = \frac{1}{2l_s^2}\: (D-1)(D-2),
\end{equation}
this string state only depends on $\alpha '$ (and $\hbar $, c). 
Size, surface gravity and temperature of the de Sitter string 
phase are 
\begin{equation}
H_s = \frac{c}{l_s}, \quad \mathcal{K}_s = \frac{c^2}{l_s}, \quad 
T_s = \frac{\hbar c}{2\pi k_B}\: \frac{1}{l_s} ,
\end{equation}
corresponding to a de Sitter maximal curvature
\begin{equation}
R_s = D \: (D-1)\: \frac{c}{l_s^2}
\end{equation}
This is also supported by the string back reaction computation \cite{1}: the 
leading term of the de Sitter curvature in the quantum regime is given by eq (4.31) 
plus 
negative corrections in an expansion in powers of ($R_{sem}/R_s$), $R_{sem}$ being the semiclassical (QFT)
de Sitter curvature. The two phases: semiclassical and stringy are dual of each other (in the sense of the classical/quantum gravity 
duality) satisfying eqs (2.21)-(2.23).  
\subsection{Quantum de Sitter spectrum}
From the asymptotic degeneracy of states $d_N$ and eq (56) we can write the 
corrections to the semiclassical entropy of de Sitter space :    
\begin{equation}
S_{sem}=S_{sem}^{(0)}-ak_B\log \left( \frac{S_{sem}^{(0)}}{k_B}\right), \quad 
S_{sem}^{(0)}=\frac{\pi k_B}{l_{Pl}}\left( \frac{c}{H}\right) ^2
\end{equation}
Similarly to the BH case,this gives us the quantization condition of de Sitter background
 \begin{equation}
S_{sem} = 2\pi k_B 2b^2\, N-ak_B \log (2\pi 2b^2 \, N) 
\end{equation}   
\begin{equation}
M_{cl}=\left(\frac{2\pi k_B}{c^2}\: T_{sem}\right) (2b)^2\, N, \quad 
b = \sqrt{\frac{D-2}{6}}
\end{equation}
\begin{equation}\mbox{that~is,} \;
M_{cl} = m_{Pl}\: 2b\sqrt{N}, \quad 
L_{cl} = l_{Pl}\: 2b\sqrt{N}
\end{equation}
Or, for $H$, $R$ and $\Lambda $ : 
\begin{equation}
H_N=\left( \frac{c}{l_{Pl}}\right)\: \frac{1}{(2b\sqrt{N})}\, , \quad 
R_N=D(D-1)\: \frac{c}{l_{Pl}^2}\: \frac{1}{(2b\sqrt{N})^2}\,,\quad
\Lambda _N=\frac{3}{4\, l_{Pl}^2}\: \frac{(D-1)(D-2)}{N} 
\end{equation}
The quantum mass de Sitter spectrum eq (4.35) is like the highly excited string 
spectrum eq (4.16), with $m_s=m_{Pl}b$. As 
($G/c^2\rightarrow \alpha '$) this spectra are the same. For increasing N 
the string levels in dS tend to mix up and pile up towards N$_{max}$, which is symptomatic feature  of a phase transition 
at the $M_s$ (or $T_s$) string temperature in de Sitter background eq (4.26). 
This phase transition is confirmed from the 
computation of the string density of states and entropy : 
from eq.(4.17) the string in de Sitter background has an entropy   
$S_s=k_B\log \rho _s$ equal to  
\begin{equation}
S_s=\frac{\pi k_B}{4}\left[ 1-\sqrt{1-\left( \frac{m}{M_s}\right) ^2}
\right] ^{1/2}+k_B\log f\left( \frac{m}{M_s}\right) 
\end{equation}
where $f(m/M_s)$ is given by eq (4.18). 
For low ($m\ll M_s$) and high ($m\rightarrow M_s$) masses we have 
\begin{equation}
S_s\,( m\ll M_s) = k_B\left( \frac{m}{m_s}\right) - ak_B\log \left( \frac{m}
{m_s}\right) 
\end{equation}
\begin{equation}
S_s\, (m\rightarrow M_s) = k_B\left( \frac{M_s}{m_s}\right) - 
ak_B\log \left( \frac{M_s}{m_s}\right) + 
k_B\log \sqrt{\frac{M_s}{m-M_s}} 
\end{equation}
Or, in terms of temperature eqs(4.26) :
\begin{equation}
S_s\, (T\ll T_s)=k_B\left( \frac{T}{T_s}\right) -ak_B\log 
\left( \frac{T}{T_s}\right) 
\end{equation}
\begin{equation}
S_s\, (T\rightarrow T_s)=k_B\left( \frac{T_s}{t_s}\right) -ak_B\log
\left( \frac{T_s}{t_s}\right) +k_B\log \sqrt{\frac{T_s}{T-T_s}} 
\end{equation}
\\
The term $\log\sqrt{\frac{M_s}{m-M_s}} = 
\log \sqrt{\frac{T_s}{T-T_s}} $ indicates the presence of a phase 
transition at the string de Sitter temperature  
$T_s = \frac{c^2}{2\pi k_B}\: (\alpha ' H)^{-1}$. 
Remarkably, the quantum mass spectrum, density of states and 
entropy of strings in de Sitter background account for the quantization (quantum mass spectrum)  of the semiclassical de Sitter  background, density of states and 
entropy and in the string regime a phase transition takes place.
The characteristic behaviour associated to this phase transition is the logartihmic singularity at $T_{s}$ (or pole singularity in the specific heat) which is the same in all dimensions. The square root branch 
point at $T_s$ is the same in all space-time dimensions. The statistical mechanics of 
self-gravitating non relativistic systems also shows a square root behaviour in the
physical magnitudes near the critical point (found by mean field and 
Monte Carlo computations)\cite{8}. This gravitational phase transition, due to the long range attractive gravitational interaction at finite temperature,  
is linked, although with more complex structure, to the Jeans'like instability, here at the string de Sitter temperature 
$T_s$ eq (4.26). 

\section{Semiclassical (QFT) and Quantum (String) Anti-de Sitter states}
AdS background alone has no event horizon and the surface gravity is zero.The Hawking temperature is zero in AdS, ie the AdS QFT vacuum has no 
intrinsic temperature,  
\begin{equation}
T_{sem\, AdS} = 0
\end{equation}
On the other hand, there is no maximal temperature for strings in AdS,\cite{5}. 
The mass spectrum of quantum strings does not have any maximal or critical 
mass and the partition function for a gas of strings in AdS is defined 
for {\it all} temperature \cite{5},\cite{6}, 
\begin{equation}
T_{s\, AdS} = \infty
\end{equation}
Again, we see that QFT and quantum strings satisfy the dual relations 
eqs (2.21)-(2.22). The (H, R, $\Lambda $) relation in AdS is the same as 
in de-Sitter space, (but R and $\Lambda $ being negative) : 
\begin{equation}
R=-D(D-1)\, \frac{H^2}{c},\quad H=c\, \sqrt{\frac{2|\Lambda |}{(D-1)(D-2)}}
\end{equation}
The quantities ($L_{cl}, \mathcal{K}_{cl}, M_{cl}, M_{sem}, T_{sem}$) 
are the same as eqs(4.8) for de Sitter background. Although there is no 
horizon, 
these quantities set up typical classical and semiclassical scales for 
length, acceleration, mass and temperature in AdS background. Similarly,  
($L_s, \mathcal{K}_s, M_s, T_s$) are typical length, acceleration, 
mass and temperature scales for strings in AdS. In particular, $M_s$ is 
{\it not} a maximal mass for strings in AdS, 
but it sets up the typical AdS string mass scale : 
(i) low, 
(ii) intermediate and (iii) high string mass states in AdS 
correspond respectively to 
\begin{equation}
(i)\: m\ll M_s,\quad (ii)\: m\sim M_s,\quad (iii)\: m\gg M_s, \quad 
\left( M_s=\frac{c}{\alpha ' H}\right), 
\end{equation}
The regime (iii) is absent in de-Sitter background.  
The mass formula for strings in AdS spacetime is given by\cite{5} 
\begin{equation}
\left( \frac{m}{m_s}\right) ^2=2\sum  _{n>0}\left( \frac{n^2+\omega _n^2}
{\omega _n}\right) +\sum _{n>0}\mathcal{N}\left( \frac{n^2+\omega _n^2}
{\omega _n}\right) 
\end{equation}
where $\mathcal{N}$ is the number-operator eq (4.13) 
and $\omega _n$ the frequency of the AdS string oscillators : $
\omega _n=\sqrt{n^2+\left( \frac{m}{M_s}\right) ^2}$.

This is the same mass formula as in de Sitter background but with positive 
sign inside the square root in  
$\omega_n$, (instabilities {\it do not} appear in AdS alone, nor for strings 
nor for QFT). 
In the conformal invariant AdS background (WZWN model), the mass formula is 
very similar \cite{7} : 
\begin{equation}
\left( \frac{n}{m_s}\right) ^2=\left( \frac{m_+}{m_s}\right) ^2 
+ \left( \frac{m_-}{m_s}\right) ^2,\quad m_s=\sqrt{\frac{\hbar }{\alpha ' c}}
\end{equation}
with
\begin{equation}
\left( \frac{m_{\pm}}{m_s}\right) ^2 = 
\sum _{n>0}\frac{(n^2+\omega _{n\pm })}{\omega _{n\pm }} + 
\sum _{n>0}\mathcal{N}_{\pm }\, \frac{(n^2+\omega _{n\pm }^2)}{\omega _{n\pm }},\quad
\omega _{n\pm }=\left| n\pm \frac{m}{M_s}\right| ,\quad 
M_s=\frac{c}{\alpha ' H}
\end{equation} 
The physics described by formulae (5.5) and (5.6) is the {\it same} 
: low and high mass states 
are {\it identical}, only intermediate mass states are slightly 
different, but these differences are minor. 
Low and high mass spectrum is given by 
\begin{equation}
\left( \frac{m}{m_s}\right) ^2 (m\ll M_s)  = 
n\left[1+O\left( \frac{m_s}{M_s}\right) ^2 \right]\end{equation}
\begin {equation} 
\left( \frac{m}{m_s}\right) ^2(m\gg M_s) = 
\left( \frac{m_s}{M_s}\right) ^2\, n^2+n,\quad n\gg (M_s/m_s)^2
\end{equation}
ie
\begin{equation}
m=\frac{m_s^2}{M_s}\, n\left[ 1+\left( \frac{M_s}{m_s}\right) ^2\frac{1}
{n}\right] ^{1/2},\quad 
\frac{m_s^2}{M_s}=\frac{1}{\alpha '}\left( \frac{l_s^2}{L_{cl}}\right) = 
\frac{\hbar H}{c^2}
\end{equation}
 the high masses  
are independent of $\alpha '$, (the scale for the ($m\gg M_s$) states is sets 
up by H (and $\hbar /c$))\cite{5},\cite{6},\cite{7} and masses increase with n (not with 
$\sqrt{n}$).
\subsection{String density of states and entropy in AdS.}
Let us derive now the density of mass levels $\rho_s (m)$. From eq (5.5) and  
the identity $\rho_s (m)\, dm=d_n(m)\, dn $,we have
\begin{equation}
\rho_s (m)=\frac{2m}{m_s^2}\: 
\frac{d_n(m)}{\left[ 1+2n\left( \frac{m_s}{M_s}\right) ^2\right]}
\end{equation}
From $d_n$(n) eq (3.23) we find : 
\begin{equation} 
\rho_s (m)=\frac{2(m/m_s^2)}{\sqrt{1+4(m/M_s)^2}}\: (2\pi b\sqrt{n})^{-a}
\: e^{2\pi b\sqrt{n}}
\end{equation}
where
\begin{equation}
2\pi b\sqrt{n}=\sqrt{2}\pi b\left( \frac{M_s}{m_s}\right) 
\left[ -1+\sqrt{1+4\left( \frac{m}{M_s}\right) ^2}\right] ^{1/2},\quad
\frac{M_s}{m_s}=\frac{L_{cl}}{l_s}=\frac{c}{H}\sqrt{\frac{c}{\alpha '\hbar}}
\end{equation}
For low, intermediate and high masses we have : 
\begin{equation}
\rho_s (m\ll M_s)=2\left( \frac{m}{m_s^2}\right)\left( 2\pi b\frac{m}{m_s}
\right) ^{-a}\: e^{2\pi b\frac{m}{m_s}\left[ 1+\left( \frac{m}{M_s}\right) 
^2+O\left( \frac{m}{M_s}\right) ^4\right] }
\left[ 1-2\left(\frac{m}{M_s}\right) ^2\right] 
\end{equation}
\begin{equation}
\rho_s (m\sim M_s)\sim 2\left( \frac{M_s}{m_s^2}\right) \left[ 2\pi b\: \left( 
\frac{M_s}{m_s}\right) \right] ^{-a}\: e^{\sqrt{2}\pi b\:  \left( \frac{M_s}{m_s}\right) }
\end{equation}
\begin{equation}
\rho_s (m\gg  M_s)=\left( \frac{M_s}{m_s^2}\right) \left( 2\pi b\frac{\sqrt{mM_s}}{m_s}\right) ^{-a}\: e^{2\pi b\frac{\sqrt{mM_s}}{m_s}\left[ 1-\frac{1}{2}
\left( \frac{M_s}{m}\right) \right]} \left[ 1+O\left( \frac{M_s}{m}\right) 
\right] 
\end{equation} 
\\
Or, in terms of temperature :  
\begin{equation}
\rho_s\, (T\ll T_s)=2\left( \frac{c^2}{2\pi k_Bt_s}\right) \left( \frac{T}{T_s}
\right) \left( 2\pi b\frac{T}{t_s}\right) ^{-a}\: 
e^{2\pi b\: T/t_s}
\end{equation}
\begin{equation}
\rho_s \, (T\sim T_s)=2\left( \frac{c^2}{2\pi k_Bt_s}\right) \left( \frac
{T_s}{t_s}\right) \left( 2\pi b\frac{T_s}{t_s}\right) ^{-a}\: 
e^{2\pi b\: T_s/t_s}
\end{equation}
\begin{equation}
\rho_s\, (T\gg T_s)=\left( \frac{c^2}{2\pi k_Bt_s}\right) \left( \frac
{T_s}{t_s}\right) \left( 2\pi b\frac{\sqrt{TT_s}}{t_s}\right) ^{-a}\: 
e^{2\pi b\frac{\sqrt{TT_s}}{t_s}}
\end{equation}
Here we have only shown the dominant behaviours, the sub-leading terms can be 
read 
directly from eqs (5.13)-(5.14). The corresponding low, medium and high temperature 
behaviours of the entropy 
$S_{s}=k_B\, \log \rho_{s} $, are read from equations (5.17)-(5.18).
We see that the characteristic mass ratio (or temperature) in $\rho_{s} (m)$ and 
$S_{s}$  goes  from (i) $m/m_s$ (or $T/t_s$) for the low excited (or weak curvature) 
states as in flat space, to the new behaviour (iii) $\sqrt{TT_s}/t_s$ for 
the highly excited states, (passing through the intermediate mass state 
ratio (ii) $T_s/t_s$) :
\begin{equation}
(i)\: \frac{m}{m_s}\: (m\ll M_s)\: \longrightarrow (ii)\: \frac{M_s}{m_s}
\: (m\simeq M_s)\: \longrightarrow 
(iii)\: \frac{\sqrt{mM_s}}{m_s}\: (m\gg M_s)
\end{equation}
\begin{equation}
\mbox{ie,}\quad 
(i)\: \frac{T}{t_s}\: (T\ll T_s)\: \longrightarrow (ii)\: \frac{T_s}{t_s}
\: (T\simeq T_s)\: \longrightarrow 
(iii)\: \frac{\sqrt{TT_s}}{t_s}\: (T\gg T_s)
\end{equation}
The characteristic low mass scale is the fundamental string 
mass $m_s$, while for the high masses, the scale is the AdS string mass 
$M_s = c/\alpha ' H$. Interestingly enough, since 
\begin{equation}
\frac{T_s}{t_s}=\frac{t_s}{T_{sem}},
\end{equation}
$\rho_s (m)$ can be also expressed in terms of the semiclassical (QFT) 
temperature scale $T_{sem}$ : 
\begin{equation}
\rho_s (TT_{sem}\ll t_s^2)=2\left( \frac{c^2}{2\pi k_Bt_s}\right) \left( 
\frac{TT_{sem}}{t_s^2}\right) \left(  2\pi b\frac{T}{t_s}\right) ^{-a} \, 
e^{2\pi b\frac{T}{t_s}} 
\end{equation}
\begin{equation}
\rho_s (TT_{sem}\sim t_s^2)=2\frac{c^2}{2\pi k_BT_{sem}}\left( 2\pi b\frac{t_s}
{T_{sem}}\right) ^{-a}\, e^{2\pi b\frac{t_s}{T_{sem}}}
\end{equation}
\begin{equation}
\rho_s (TT_{sem}\gg t_s^2)=2\frac{c^2}{2\pi k_BT_{sem}}\left( 2\pi b\sqrt{
\frac{T}{T_{sem}}}\right) ^{-a}\, e^{2\pi b\sqrt{\frac{T}{T_{sem}}}}
\end{equation}
$t_s\, (\alpha ' )$ present in the low excited regime (i) eq (5.23),   
{\it completely cancels out} in the highly excited regime (iii) eq (5.25). 
Eqs (5.17) or eqs (5.23), show that the low excited AdS string 
state (low values of $\Lambda $, low curvature), behaves like strings in 
flat space. 
For this state, $\rho_s (m)$, entropy, Hagedorn temperature take the flat 
space values, multiplied by corrections 
$\left[ 1+\sum\limits^\infty _{n=1}c_n\;\left(\frac{\alpha ' H^2}{c}\right) ^n
\right]$. The highly excited AdS state (high values of $\Lambda $, high 
curvature) is {\it very different} from flat space : the 
level spacing increases with n; $\rho_s (m)$, entropy, etc, are functions of 
$\sqrt{T/T_{sem}}$, there is {\it no} maximal temperature, there is 
{\it no} finite (at finite temperature) critical point : it is pushed 
up to infinity by the negative $\Lambda $. There is no singularity at 
$T\rightarrow T_s$ in $\rho_s (m)$ nor in the entropy as in flat space or as in 
de Sitter space. In AdS, 
$\rho_s (m)$, the partition function, etc, are {\it all finite}, 
no phase transition occurs in AdS alone.\\
Notice that the so called Hawking-Page phase transition \cite{9} which appears 
in  black hole-AdS systems (BH-AdS) in the context of QFT is due to the black hole (not to AdS) : AdS space 
just sets up the boundary condition in asymptotic space such that the 
system BH-AdS be thermodynamically stable, and the semiclassical BH undergoes a transition 
at a critical temperature into a thermal gas. (BH's in dS space should also undergo a similar transition, although they are 
not, of course, thermally stable. Such phase transition should be of the same type 
of that found in the thermodynamics of self gravitating systems \cite{8}).  
\subsection{Quantum Anti-de Sitter spectrum}
The role played by $M_s$ (or $T_s$ scale) in AdS is clear : $M_s$ is 
the mass from which AdS differs drastically from flat space. For 
$T\sim T_s$ ($m\sim M_s$) or for $T\gg T_s$ ($m\gg M_s$), the string mass 
spectrum or $\rho_s (m)$ in AdS gives us a quantization condition for $M_s$ 
(or H). Since 
\begin{equation}
\frac{T_s}{t_s}=\frac{M_s}{m_s}=\frac{L_{cl}}{l_s}=\sqrt{\frac{c}{\alpha ' 
\hbar }}\: \frac{c}{H}
\end{equation} 
from $\rho_s (m)$ eq (5.18), we have at leading order 
\begin{equation}
M_{sn}=m_s\sqrt{n},\quad \mbox{ie}\quad 
H_n=\frac{c}{l_s}\: \frac{1}{\sqrt{n}},\quad \Lambda _n = \frac{1}{2l_s^2}\, 
(D-1)(D-2)\, \frac{1}{n},
\end{equation}
which is the quantum spectrum for the background.
\\ \\
For $T\gg T_s$, $\rho_s (m)$ eq (5.19) leads 
\begin{equation}
\sqrt{\frac{T}{T_{sem}}}=\sqrt{\frac{m}{M_{sem}}}=\sqrt{n},\quad \mbox{ie}
\quad m\: M_{cl}=m_{Pl}^2\: n,
\end{equation}
(which is the high n string spectrum in AdS). And when 
$m\rightarrow M_{cl}$, this yields 
\begin{equation}
M_{cln} = m_{Pl}\: \sqrt{n},\quad \mbox{ie}\quad 
H_n = \frac{c}{l_{Pl}}\: \frac{1}{\sqrt{n}}\: ,
\end{equation}
which is the quantization of $M_{cl}$ (or H) eq (5.27) (with $l_{Pl}$ 
instead of $l_s$, ie $G/c^2\rightarrow \alpha ' $). Again, when $ m\sim
M_s $ and/or when $ m\gg M_s$ , the string becomes the background (and conversely,
the high mass string spectrum accounts for the quantum spectrum of the
background (the background becomes the string)).   
\section{Concluding Remarks}

We have provided new results and understanding to the conceptual unification of the quantum properties of BH's, elementary particles, dS and AdS states. This is achieved by recognizing the relevant scales of the semiclassical (QFT) and quantum (string) regimes of gravity. They turn out to be the classical-quantum duals of each other, in the precise sense of the wave-particle (de Broglie, Compton) duality, here extended to the quantum gravity (Planck) domain: wave-particle-string duality.
\\
Concepts as the Hawking temperature and string (Hagedorn) temperature are shown to be precisely the same concept in the different (semiclassical and quantum) gravity regimes respectively. Similarly, it holds for the Bekenstein-Hawking  entropy  and string entropy. An unifying formula for the density of mass states and entropy of BH, dS and AdS states has been provided in the two: semiclassical and string regimes. Quantum and string properties of the backgrounds themselves have been extracted. This is particularly enlighting for de Sitter background for several reasons:
\\
(i) the physical (cosmological) relevance of dS background (inflation at the early time, present acceleration);
\\
(ii) the lack at the present time of a full conformal invariant dS background in string theory; 
\\
(iii) the results of string dynamics in conformal and non conformal string backgrounds which show that the physics remain mainly the same in the two class of backgrounds (conformal and non conformal) \cite {5}, \cite{6}, \cite{7},\cite{28}; 
\\
(iv) the phase transition found here for strings in dS background at the dS string temperature $T_{s}$ eq(4.26), this temperature is the precise quantum dual of the semiclassical (QFT Hawking-Gibbons) dS temperature eq(4.2). This phase transition for strings in dS is the analogue of the Carlitz phase transition \cite {29} for the thermodynamics of strings in flat space and in BH's \cite{1}.
\\ 
A precise picture relating these string phase transitions among themselves and to the classical and semiclassical (QFT) gravitational 
phase transitions desserves future investigation but starts to be outlined here:
\\ 
(a) the string phase transitions at the string temperature $T_{s}$ found in Minkowski \cite {29}, BH \cite {1}, and dS backgrounds [this paper] are all of the same nature (the value of $T_{s}$ is different in each case). 
\\
(b) This string phase transition does not occur in AdS alone (ie without BH,s), it could occurs in BH-AdS backgrounds. 
\\
(c) These phase transitions are the string analogues of the so-called Hawking-Page transition \cite{9} occuring in the semiclassical gravity regime (and which is due to the BH). Hawking-Page transition does not occurs in AdS alone (ie, without the BH).
\\
(d) All these phase transitions (for strings and QFT) should be the counterparts of the classical gravitational phase transitions occuring in the statistical mechanics of the selfgravitating gaz of particles \cite{8} whose origin is (non linear) Jean's instability at finite temperature (usually called ``gravothermal instability'' \cite{37}, but with a richer structure: critical points, metastable phases, inhomogeneous ground state, fractal dimension \cite {8}.
\\ \\
The description of the last stages of BH evaporation is not the central issue of this paper, but our results show that BH evaporation ends as quantum string decay into elementary particle states (most massless), ie pure (non mixed) quantum states.
\\ 
The effects of BH angular momentum and charges can be included in this framework, as well as those of a cosmological constant to consider BH-dS , BH-AdS, etc , although we do not treat them here.
\\ \\   
The exhibit of $(\hbar, c, G)$ or $(\hbar, c, \alpha')$ helps in recognizing the different relevant scales and regimes. Even if a hypothetical underlying ``theory of everything'' could only 
require pure numbers (option three in \cite {30}), physical touch at some level asks for the use of fundamental constants \cite {31},\cite{30},\cite{32},\cite {33}. Here we favour {\it three} fundamental constants \cite {31},\cite {32},\cite{33} $((\hbar, c, G)$, or $(\hbar,c,\alpha'))$, (tension being $c^2/G$ or $1/\alpha'$). (We do not use any further coupling, conceptual results here will not change by further couplings or interactions), (Hawking radiation, intrinsic gravitational entropy and other thermal features do not change by interactions or further background fields, dilaton, etc)
\\ \\
This paper does not make use of conjectures, proposals, principles, (CPP's) formulated in the last years in connection with string theory \cite {34},\cite {14}. String dynamics in curved backgrounds started well before CPP's,\cite{23}, many results for strings in curved backgrounds addressed nowdays (strings on AdS, on plane waves, semiclassical and canonical string quantization, rotating strings, null strings, integrable string equations, time dependent string backgrounds) can be found in \cite {35}, \cite {36} and refs therein.

\end{document}